\documentclass[twocolumn,showpacs,amsmath,amssymb,prd]{revtex4}
\usepackage{graphicx}
\usepackage{dcolumn}
\usepackage{bm}
\usepackage{epsfig}
\usepackage[a4paper]{geometry}
\usepackage[in]{fullpage}
\usepackage{color}
\newcommand{\beq}{\begin{equation}}
\newcommand{\eeq}{\end{equation}}
\newcommand{\beqarray}{\begin{eqnarray}}
\newcommand{\eeqarray}{\end{eqnarray}}

\begin{document}

\title{A lepto-hadronic model of gamma rays from the Eta Carinae and prospects for neutrino telescopes}


\author{Nayantara Gupta$^1$}
\email{nayan@rri.res.in}

\author{Soebur Razzaque$^2$}
\email{srazzaque@uj.ac.za}

\affiliation{$^{1}$Raman Research Institute, C.V.\ Raman Avenue, Sadashivanagar, Bangalore 560080, India\\
$^{2}$Department of Physics, University of Johannesburg, P.O.\ Box 524, Auckland Park 2006, South Africa}

\begin{abstract}
The stellar binary $\eta$ Carinae has been observed during its full orbital period in gamma rays by the Fermi-{\it Large Area Telescope} (LAT).  The shock-accelerated electrons in the colliding winds of the two stars radiate synchrotron photons in the magnetic field of the shocked region and inverse Compton photons, where the target photons are from the thermal emissions by the more massive and luminous of the two stars.  The inverse Compton emission dominates the gamma-ray flux data from the $\eta$ Carinae, however the spectral energy distribution shows signature of a hadronic component in the $\sim 10$-300~GeV range during the periastron passage. Current and future air Cherenkov telescopes will be able to constrain this component at TeV energies. Acceleration of cosmic-ray protons to $\gg 1$~TeV energies in the colliding winds, required to explain the hadronic emission component through photopion interactions,  can lead to detectable signal of $\gtrsim 10$~TeV neutrino  events in large kilometer scale neutrino telescopes. 
\end{abstract}

\pacs{95.85.Ry, 98.70.Sa}
\date{\today}
\maketitle

\section{Introduction}
$\eta$ Carinae is one of the most massive and brightest stellar binaries in our Galaxy, located only at a distance of 2.3~kpc.  The heavier star  $\eta$ Car-A, of mass in the range of 80-$120\,M_\odot$ and the lighter star $\eta$ Car-B, of mass less than $30\,M_\odot$, are rotating around each other in  2022.7$\pm$1.3 days~\cite{Nielsen:2007ht,Damineli:2007tc}.  It was known to the past astronomers for its series of giant outbursts in 1840s and 1890s~\cite{Frew:2004}.   The historical outbursts resulted in ejection of fast moving gas and subsequently the bipolar nebula known as Homunculus was formed~\cite{Frew:2004}.  A blast wave moving with a speed of 3500 to 6000~km~s$^{-1}$ surrounds  the binary system which originated in the great eruption in 1843.  Colliding stellar wind from both the stars may produce temperature as high as $10^7$~K in the shock heated region which results in emission of thermal X-rays.  The X-ray light curve of $\eta$ Carinae shows orbital variation and has been modeled in~\cite{Hamaguchi:2007iv,Parkin:2009fh, Bednarek:2011yn}.  

Detection of GeV gamma rays from the $\eta$ Carinae region with the {\it Astro-Rivelatore Gamma a Immagini Leggero} (AGILE)~\cite{Tavani:2009wm} and the Fermi-LAT~\cite{collaboration:2010ss} clearly requires acceleration of particles to highly-relativistic energies in a shock-acceleration process.  Such a possibility has been discussed for stellar binaries with colliding wind in literature~\cite{Eichler:1993,Reimer:2005hy,Bednarek:2011yn}.  Most recently Fermi-LAT detected 0.2-300~GeV gamma rays from the full orbit of the binary system~\cite{Reitberger:2015fha,Balbo:2017vum}.  In particular, a mild orbital modulation in gamma-rays has been reported with the spectrum during the periastron phase showing a hardening above $\approx 10$~GeV while no such hardening was detected during the apastron phase.

Multi-wavelength data from $\eta$ Carinae have been modeled earlier with radiative emissions by shock accelerated electrons and protons in colliding wind from the binary stars. The electrons lose energy by synchrotron emission and, more importantly, by inverse Compton (IC) losses through the thermal photons of temperature tens of thousands of Kelvin emitted by $\eta$ Car-A, which has a luminosity of $5\times 10^6 L_{\odot}$~\cite{Madura:2011gu}.  While IC emission alone fits gamma-ray spectrum during the apastron phase, the periastron phase spectrum requires a hadronic component at $\gtrsim 10$~GeV, possibly from cosmic-ray proton interactions with wind particles and resulting in neutral pion decay~\cite{far}. The gamma ray emission is dominantly  from hadronic interactions in the model discussed in~\cite{Ohm:2015uha}. 

In this paper, we have fitted multi-wavelength spectral data including average gamma-ray spectra recorded over 500 days centered on the periastron and apastron phases~\cite{Reitberger:2015fha} using a combined leptonic-hadronic model of IC and $\pi^0$-decay emissions.  We have also calculated neutrino flux expected from hadronic interactions in the colliding winds of the binaries and explored prospects for detecting neutrino events with the IceCube~\cite{Achterberg:2006md} and KM3NeT~\cite{Adrian-Martinez:2016fdl} neutrino telescopes. 

The organization of this paper is following.  In Section II we discuss properties of the binary system, colliding winds, shocks and particle acceleration.  In Section III we detail our modeling of spectra, focusing on gamma-ray data.  We describe our neutrino flux model in Section IV and discuss detection prospects by neutrino telescopes.  We compare our results with earlier works and conclude in Section V.

\section{Colliding winds and shocks}
Stars of the $\eta$ Carinae system blow stellar wind with mass-loss rates of ${\dot M}_{w,\rm A} =10^{-3}M_{\odot}$~yr$^{-1}$ and ${\dot M}_{w,\rm B} =10^{-5}M_{\odot}~$yr$^{-1}$ and speeds $v_{w,\rm A}=500$~km~s$^{-1}$ and $v_{w,\rm B} = 3000$~ km~s$^{-1}$, respectively for $\eta$ Car-A and B~\cite{pit02}.  A shock region is formed at an intermediate stagnation position between the two stars where wind pressures are equal.  The pressure $P_w$ and density $\rho_w$ of stellar wind at a radius $R$ from the star is
\begin{equation}
P_w=\frac{1}{2}\rho_w v_w^2= \frac{{\dot M}_w v_w}{8\pi R^2}\,.
\end{equation}
The pressure balance gives the ratio of the distances to the stagnation position, where shocks are formed, from the two stars $\eta$ Car-A and B as
\begin{equation}
\beta = \frac{R_{\rm s,A}}{R_{\rm s,B}} = \sqrt{\frac{{\dot M}_{w,\rm A} v_{w,\rm A}}{{\dot M}_{w,\rm B} v_{w,\rm B}}} = 4.1\,.
\label{ratio}
\end{equation}

We adopt a description of the $\eta$ Carinae orbit with $\eta$ Car-A at one of the focii of an ellipse~\cite{Pittard:1998} with semi major axis $a=15.4$~AU and orbital eccentricity $e=0.9$~\cite{Madura:2011gu}.  The distance between the stars for any given angle $\phi$, measured from the major axis to the position of $\eta$ Car-B, can be found by solving a parametric equation for an ellipse as
\begin{equation}
R_{\rm s,A} + R_{\rm s,B} = \sqrt{(ea-a\cos{\phi})^2 + (1-e^2) a^2 \sin^2{\phi}}\,,
\label{dis}
\end{equation}
where the separation is 1.5~AU and 29.3~AU during the periastron ($\phi = 0$) and apastron ($\phi = \pi$) points, respectively.  Solving Eqs.~(\ref{ratio}) and (\ref{dis}) we obtain the distances as
\begin{eqnarray}
(R_{\rm s,A}; R_{\rm s,B}) = (0.8; 0.2)~{\rm AU} ~~~~~~~~~~~~~~~~~~ \nonumber \\
 ~~~~ \times \sqrt{(13.9-15.4\cos\phi)^2+45.1\sin^2{\phi}}\,.
\label{distAB} 
\end{eqnarray}
The values $R_{\rm s,A} = (1.2, 23.5)$~AU at $\phi=(0, \pi)$ are comparable to the values obtained with detailed numerical simulations of the colliding winds~\cite{Pittard:1998}.  The angle $\phi$ is related to the ``phase'' as $P = \phi/2\pi$, and $P = 0$ at the periastron and $P=0.5$ at the apastron.

The surface area of the shocked region depends on the asymptotic angle $\theta_\infty$ at which $R_{\rm s}\to\infty$, and can be found from the equation $\theta_\infty - \tan\theta_\infty = \pi/(1-\beta)$~\cite{Canto:1996}.  Using $\beta = 4.1$ from Eq.~(\ref{ratio}) we find $\theta_\infty = 65^\circ$.  The shock surface area from $\eta$ Car-A is therefore $\propto 4\pi R_{\rm s,A}^2 \sin^2(\theta_\infty/2) \propto R_{\rm s,A}^2$.  It is interesting to note here and as we will see shortly, that the radiation from particles scales as the energy density of the target radiation field or material, which scales as $\propto 1/R_{\rm s,A}^2$.  As a result, the radiative flux is constant throughout the orbit to the first approximation.  Indeed the gamma-ray flux from $\eta$ Carinae vary at most by a factor of two, with large error bars, throughout its orbital phase~\cite{Reitberger:2015fha, Balbo:2017vum}.

The wind kinetic luminosity is $L_w = {\dot M}_w v_w^2/2 \approx 8\times 10^{37}$~erg/s, which corresponds to an energy density $u_w = L_w/4 \pi R^2 v_w$.  We assume that a fraction $\epsilon_B$ of $u_w$ can be converted to random magnetic field $B$ with energy density $B^2/8\pi$ in the shock region due to plasma instabilities.  For the shock from $\eta$ Car-A, this magnetic field is
\begin{eqnarray}
B_{\rm s,A} = 5\, \Big(\frac{\epsilon_B}{0.18}\Big)^{1/2} \Big(\frac{{\dot M}_{w,\rm A}}{10^{-3}~M_{\odot}~{\rm yr}^{-1}}\Big)^{1/2} 
\nonumber \\ 
\times \Big(\frac{v_{w,\rm A}}{500~{\rm km/s}}\Big)^{1/2} \Big(\frac{R_{\rm s,A}}{10~{\rm AU}}\Big)^{-1}~{\rm G} \,.
\label{mag}
\end{eqnarray}
{\ 
We consider particle acceleration rate in perpendicular shocks which is faster than the rate in parallel shocks.  The acceleration time scale for a perpendicular shock with the magnetic field $B_{\rm s,A}$ to an energy $E$ is $t_{\rm acc}^\perp = 8Ec/3\zeta e B_{\rm s,A} v_{w,\rm A}^2$, where $\zeta \sim {\cal O}(10)$ is a multiplicative factor to the Bohm diffusion coefficient~\cite{Gaisser:1990vg, Protheroe:1998hp}.
}

The maximum electron energy is limited by IC losses as required to fit gamma-ray data, which we will discuss in the next section.  The energy density in starlight is $u_{\star} = L_{\star}/4\pi R^2 c$, and for the shock radius from $\eta$ Car-A is $u_{\star,\rm A} = (149, 0.4)$~erg/cm$^3$, respectively at the periastron and apastron phase, with $L_{\star,\rm A} = 5\times 10^6L_\odot$.  The IC cooling time in the Thomson limit is $t_{\rm IC} = 3m_e^2c^3/4\sigma_{\rm T} E_eu_{\star,\rm A}$. This time scale is much shorter than the shock crossing or advection time scale $3 R_{\rm s,A}/v_{w,\rm A}\sim (0.1-2)\times 10^7$~s, as also has been noted in~\cite{far}. From the condition $t^\perp_{\rm acc} = t_{\rm IC}$ we get the maximum electron energy as 
\begin{eqnarray}
E_{e,\rm max} &\approx & 57\, \Big(\frac{\zeta_e}{10}\Big)^{1/2} 
\Big(\frac{B_{\rm s,A}}{5~{\rm G}}\Big)^{1/2} 
\nonumber \\ && \times 
\Big(\frac{u_{\star,\rm A}}{2.3~{\rm erg/cm}^{3}}\Big)^{-1/2} ~{\rm GeV} \, ,
\label{eemax}
\end{eqnarray}
for the same parameters as in Eq.~(\ref{mag}).  Next we calculate the maximum proton energy.

Shock-accelerated protons interact with particles in the stellar wind through $pp$ interactions.  The density of particles in the stellar wind is $n_{\rm H} = \rho_w/m_p = {\dot M}_w/4\pi R^2 v_w m_p$ and the corresponding $pp$ interaction time scale is $t_{pp} = 1/n_{\rm H} \sigma_{pp} c$.  Here we have taken an average $pp$ interaction cross-section $\sigma_{pp} \approx 30$~mb.  By equating $t^\perp_{\rm acc}$ to $t_{pp}$ we calculate the maximum proton energy as
\begin{eqnarray}
E_{p,\rm max} &\approx & 1\, \Big(\frac{\zeta_p}{50}\Big) \Big(\frac{B_{\rm s,A}}{5~{\rm G}}\Big) 
\nonumber \\
&& \times
\Big(\frac{n_{\rm H, A}}{2.7\times 10^9~{\rm cm}^{-3}}\Big)^{-1} ~{\rm PeV} \, ,
\label{emax}
\end{eqnarray}
for the same parameters as in Eq.~(\ref{mag}).  The maximum proton energy obtained by equating $t^\perp_{\rm acc}$ to the shock crossing or advection time scale~\cite{Bednarek:2011yn} is larger, since $t_{pp}$ is shorter than the advection time in our case.

We note that the shock on the side of $\eta$ Car-B has similar characteristics and can be computed in the same way. However the particle density is much lower on the side of $\eta$ Car-B, with two orders of magnitude lower mass-loss rate and a factor of six faster wind.  As a result $pp$ interaction is not efficient.  The luminosity of $\eta$ Car-B is unknown and is expected to be $< 10^6L_\odot$~\cite{Madura:2011gu}.  Therefore IC emission from $\eta$ Car-B can also be smaller than from the side of $\eta$ Car-A.

\section{Modeling Spectral Energy Distributions}
\begin{figure}
\includegraphics[scale=0.41]{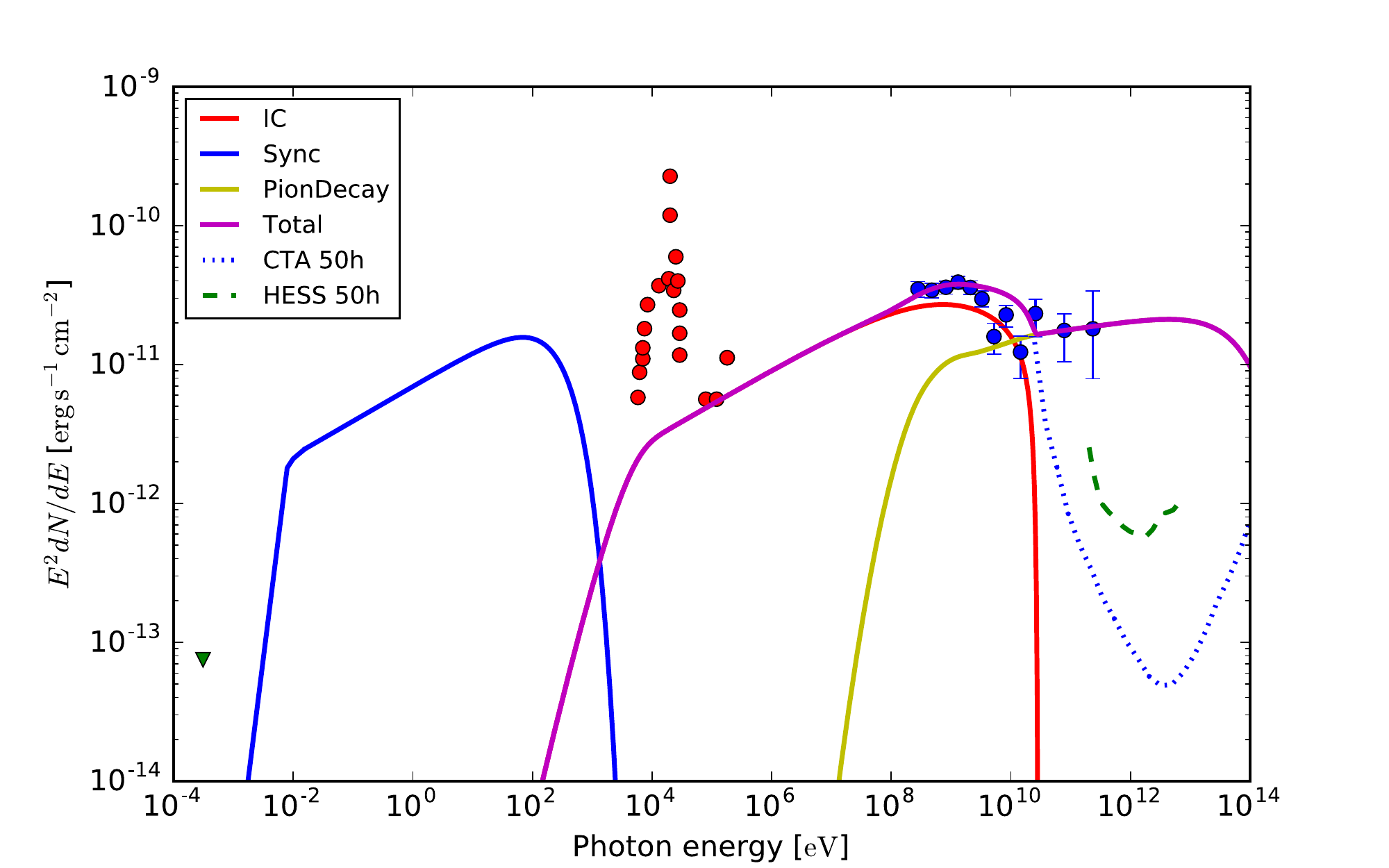}
\caption{Spectral energy distribution of $\eta$ Carinae during the 500-day periastron passage.  The radio upper limit at 3 cm (green downward triangle)~\cite{Duncan:1995} and X-ray data (red solid circles)~\cite{viotti},~\cite{leyder08},~\cite{leyder10} are plotted in addition to the gamma ray data (blue solid circles)~\cite{Reitberger:2015fha},~\cite{Balbo:2017vum}.  
The synchrotron, inverse Compton and $\pi^0$ decay gamma ray model curves are shown as solid lines.  The sum of IC and $\pi^0$ decay gamma ray is shown with the magenta line.  Also shown are 50 hour flux sensitivity curves for the CTA South and H.E.S.S. with dotted and dashed lines, respectively~\cite{Acharya:2017ttl}.}
\label{Figure--1}
\end{figure}

\begin{figure}
\includegraphics[scale=0.41]{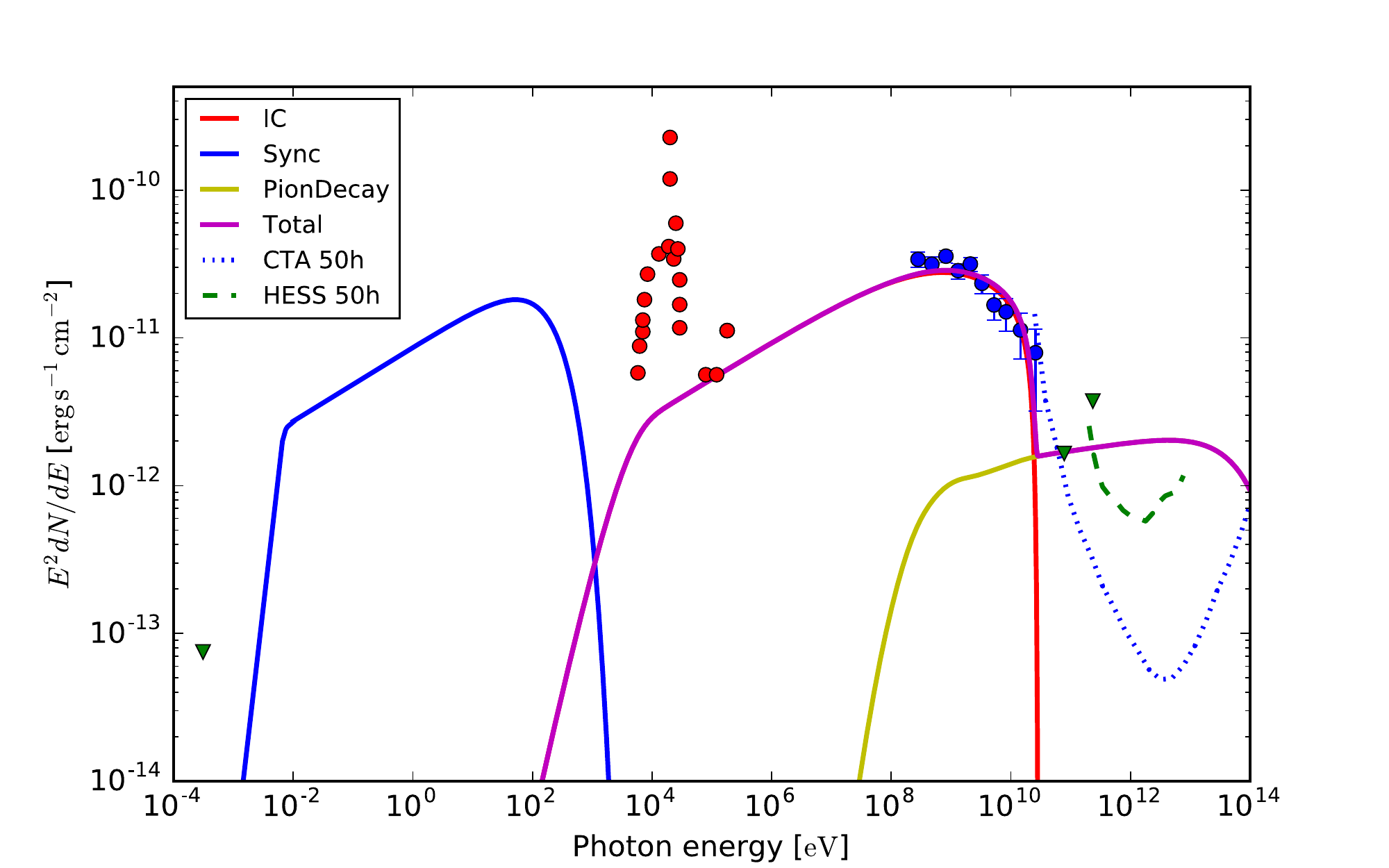}
\caption{The SED of $\eta$ Carinae during the 500-day apastron passage.  Colors of the data points and lines are the same as in Fig.~\ref{Figure--1}. The green downward triangles 
 correspond to the upper limits on the radio and gamma ray flux.}
\label{Figure--2}
\end{figure}

Fermi-LAT collected gamma-ray data from the full orbit of $\eta$ Carinae during 2024 days, from August 4, 2008 to February 18, 2014~\cite{Reitberger:2015fha}.  Further coverage, until July 1, 2015, is also reported in~\cite{Balbo:2017vum}.  Gamma-ray spectral data in the 0.2-300 GeV energy range and averaged over 500 days centered around the periastron and apastron passages are shown in Figs.~\ref{Figure--1} and \ref{Figure--2}, respectively.  Also shown in the figures are X-ray data points from BeppoSAX/MECS~\cite{viotti} and INTEGRAL/ISGRI observations~\cite{leyder08,leyder10}.  The radio upper limit in the figures at 3 cm is due to continuum emission~\cite{Duncan:1995}. 
 
Note that the phase values during the 500 day periastron and apastron passages are approximately $P=0.06$-0.91 and $P=0.33$-0.67, respectively.  To model the spectral energy distribution (SED) during these two epochs, we also average over physical quantities.  For example, during the periastron passage, the average energy density in starlight from $\eta$ Car-A is 
\begin{equation}
\langle u_{\star,\rm A} \rangle_{\rm per} = \frac{L_{\star,\rm A}}
{4\pi \langle R_{\rm s, A}^2 \rangle_{\rm per}c} = 1.7\times 10^{12}~\frac{{\rm eV}}{{\rm cm}^3}\,,
\label{starAraddensity}
\end{equation}
and the average number density of particles in the wind is
\begin{equation}
\langle n_{\rm H, A} \rangle_{\rm per} = \frac{{\dot M}_{w,\rm A}}
{4\pi \langle R_{\rm s, A}^2 \rangle_{\rm per} v_{w,\rm A} m_p c^2} = \frac{3.2\times 10^{9}}{{\rm cm}^3}\,,
\label{lumdensA}
\end{equation}
where $\langle R_{\rm s, A}^2 \rangle_{\rm per}$, with $\langle R_{\rm s,A}\rangle_{\rm per} = 1.4\times 10^{14}$~cm, is obtained from Eq.~(\ref{distAB}) after averaging over the angle $\phi$.  These densities are smaller by a factor $\langle R_{\rm s, A}^2 \rangle_{\rm per}/\langle R_{\rm s, A}^2 \rangle_{\rm aps} \approx 0.56$ during the apastron passage.

We use the publicly available NAIMA code~\cite{naima} for modeling the SEDs in Figs.~\ref{Figure--1} and \ref{Figure--2}. In this code the radiation models are implemented in modular way. Our input parameters are the strength of the magnetic field in the shock region of $\eta$ Car-A, temperature and energy density of the radiation field of $\eta$ Car-A, minimum and maximum energies of electrons and protons, and the average number density of particles in wind.  The normalization constants and the spectral indices of the relativistic electron and proton spectra in the shock region are kept as free parameters.  They are adjusted to fit the observed spectral energy distributions of photons during the periastron and apastron phases.

The dominant contribution to the GeV emission comes from IC emission by electrons, while synchrotron losses is rather small. {\ In Fig.~\ref{Figure--1} the magnetic field is set at 5~G, which correspond to $\epsilon_B = 0.16$ in Eq.~(\ref{mag}) for $\langle R_{\rm s,A}\rangle_{\rm per} = 9.4$~AU.} We kept the energy density $\langle u_{\star,\rm A} \rangle_{\rm per}$ for IC emission and number density $\langle n_{\rm H, A} \rangle_{\rm per}$ for $\pi^0$ emission fixed and varied the electron and proton spectra until getting a good fit to data.  The electron spectrum used to fit the data is
\begin{equation}
\frac{dN_e(E_e)}{dE_e}= 1.0 \times 10^{40}
\Big(\frac{E_e({\rm eV})}{511\times 10^3}\Big)^{-2.5} {\rm eV}^{-1}\,,
\end{equation}
in the energy range of 5.11~MeV to 30~GeV for the periastron passage (Fig.~\ref{Figure--1}).  {This maximum energy requires $\zeta_e = 3$ in Eq.~(\ref{eemax}) with the average radiation density in Eq.~(\ref{starAraddensity}).  The synchrotron self-absorption photon energy is $7.6\times 10^{-3}$~eV, which is higher than the frequency of the upper limit on radio flux in Fig.~\ref{Figure--1}.  The total energy required in electrons is $W_e= 2.7\times 10^{39}$ erg.}

We used a proton spectrum during the periastron passage as
\begin{equation}
\frac{dN_p(E_p)}{dE_p}= 7 \times 10^{33} 
\Big(\frac{E_p({\rm eV})}{10^9}\Big)^{-2} {\rm eV}^{-1} \,,
\label{pspecper}
\end{equation}
in the energy range of 1.22~GeV to 1~PeV. {For the particle density in Eq.~(\ref{lumdensA}) and 5~G magnetic field, attaining $E_{p,\rm max} = 1$~PeV requires $\zeta_p = 50$ in Eq.~(\ref{emax}).  The total energy required in protons is $W_p=1.6\times 10^{41}$ erg, and the ratio $W_e/W_p \approx 0.02$.}  We can compare these energies with the total energy in wind, $W_w = 8.5\times 10^{43}$~erg, which we obtain by multiplying the wind kinetic luminosity $L_w$ and the advection time scale.

A difference in gamma-ray flux between two periastron passages has been pointed out in~\cite{Balbo:2017vum}.  In particular the flux during the second periastron passage is somewhat lower.  A change in mass-loss by the stars could be a plausible reason.

For the apastron passage (Fig.~\ref{Figure--2}), we used the average densities $\langle u_{\star,\rm A} \rangle_{\rm aps} = 0.56\langle u_{\star,\rm A} \rangle_{\rm per}$ for IC emission and $\langle n_{\rm H, A} \rangle_{\rm aps} = 0.56\langle n_{\rm H, A} \rangle_{\rm per}$ for $\pi^0$ emission.  In this case, the entire GeV data can be modeled with IC emission.  The upper limit on the Fermi LAT data constrains the hadronic component of gamma rays to a lower value during the apastron passage compared to the periastron passage.  
The electron and proton fluxes required to fit data in this case are
\begin{equation}
\frac{dN_e(E_e)}{dE_e}= 2.0\times 10^{40}
\Big(\frac{E_e({\rm eV})}{511\times 10^3}\Big)^{-2.5} {\rm eV}^{-1}\,,
\end{equation}
in the 5.11~MeV-30~GeV energy range, and
\begin{equation}
\frac{dN_p(E_p)}{dE_p}\lesssim 1.2\times 10^{33} 
\Big(\frac{E_p({\rm eV})}{10^9}\Big)^{-2} {\rm eV}^{-1} \,,
\end{equation}
respectively.   {The magnetic field in this case is 3.8~G, with the same parameter value $\epsilon_B = 0.16$ in Eq.~(\ref{mag}).   The parameter $\zeta_e = 3$ in Eqs.~(\ref{eemax}) is also the same as in the periastron passage. The synchrotron self-absorption photon energy in this case is $6.6\times 10^{-3}$~eV.  The total energy in electrons and protons during the apastron passage is $W_e = 5.3\times 10^{39}$~erg and $W_p \lesssim 2.7\times 10^{40}$~erg, respectively.}   We note that the electron spectrum during the apastron passage is roughly a factor of two higher than the spectrum during the periastron passage, which compensates for the roughly a factor of two lower energy density in starlight in the former case. 

We have also plotted 50 hour flux sensitivity curves for the {\it High Energy Stereoscopic System} (H.E.S.S.) and the upcoming  {\it Cherenkov Telescope Array} (CTA) Southern site in Figs~\ref{Figure--1} and \ref{Figure--2} with dashed and dotted lines, respectively~\cite{Acharya:2017ttl}.  Note that  in near term H.E.S.S. has very good chance to detect $\gtrsim 100$~GeV gamma rays from $\eta$ Carinae before the next periastron passage in 2020.  CTA will be able to constrain if there is a cutoff in the gamma-ray spectra and shed lights on differences between $\gtrsim 30$~GeV gamma-ray emission at different orbital periods.

\subsection{$\gamma\gamma$ pair production}
High-energy gamma rays produced in the shocked-region of the colliding winds may interact with stellar photons to produce electron-positron pairs~\cite{Reimer:2005hy, Bednarek:2014esa, Ohm:2015uha}, and thus absorbed~\footnote{The electron-positron pairs, however, can Compton scatter stellar photons again to produce a secondary cascade emission.}.  This affects gamma rays with energy $2m_ec^2/kT(1+\cos\psi) \sim 100/(1+\cos\psi)$~GeV, where the temperature of $\eta$ Car-A is  assumed to be $T=60,000$~K and $\psi$ is the angle between the gamma ray and target stellar photon directions.  

To calculate the $\gamma\gamma$ opacity for gamma rays originating from the shock radius $\langle R_{\rm s,A} \rangle$ with the stellar photons from $\eta$ Car-A we follow the formalism in~\cite{Dubus:2005cw} for the case when the distance of interaction $d$ from the star is much larger than the stellar radius. 

The differential density of stellar photons per unit solid angle from $\eta$ Car-A at the shock radius $\langle R_{\rm s,A} \rangle$ is given by
\begin{equation}
\frac{dn}{d\epsilon d\Omega} = \frac{\kappa}{4\pi^3 (\hbar c)^3} \frac{\epsilon^2}{e^{\epsilon/kT} - 1}\,,
\end{equation}
where $\kappa$ is a renormalization factor defined, e.g., during the periastron phase as $\kappa = \langle u_{\star,\rm A} \rangle_{\rm per}/a_BT^4$, where $a_B$ is the Stefan-Boltzmann constant for black body energy density.  The $\gamma\gamma$ opacity for gamma rays produced in the shock and propagating to an observer is then given by
\begin{eqnarray}
\frac{d\tau_{\gamma\gamma}}{dl} &=& 
\frac{\kappa}{4\pi^2 (\hbar c)^3} \frac{\langle R_{\rm s,A}^2\rangle}{d^2} 
\nonumber \\ && \times 
\int_{\epsilon_{\rm min}}^{\infty} 
\frac{\epsilon^2 \sigma_{\gamma\gamma} (1+\cos\psi)}{e^{\epsilon/kT} -1} d\epsilon \,,
\label{opacity}
\end{eqnarray}   
where $\sigma_{\gamma\gamma}$ is the cross section~\cite{Gould:1967zzb}, $\epsilon_{\rm min} = 2m_e^2c^4/E_\gamma (1+\cos\psi)$ is the threshold energy. Here $d^2 = \langle R_{\rm s,A}^2\rangle + l^2 - 2 \langle R_{\rm s,A} \rangle l \cos\psi_0$, with $\psi_0$ is an angle the gamma ray makes with the radial direction from the star~\cite{Dubus:2005cw}.  Note that the angle $\psi$ in Eq.~(\ref{opacity}) is related to $l$, $\langle R_{\rm s,A} \rangle$ and $\psi_0$ as~\cite{Dubus:2005cw}
\begin{equation}
\psi = \tan^{-1} \left( 
\frac{\langle R_{\rm s,A}\rangle \sin\psi_0}{\langle R_{\rm s,A}\rangle \cos\psi_0 - l}
\right) \, ; \, l < \langle R_{\rm s,A}\rangle \cos\psi_0 \,.
\end{equation}
For $l > \langle R_{\rm s,A}\rangle \cos\psi_0$, the right hand side of the above expression is added with $\pi$.

\begin{figure}
\vskip 0.4cm
\includegraphics[scale=0.37]{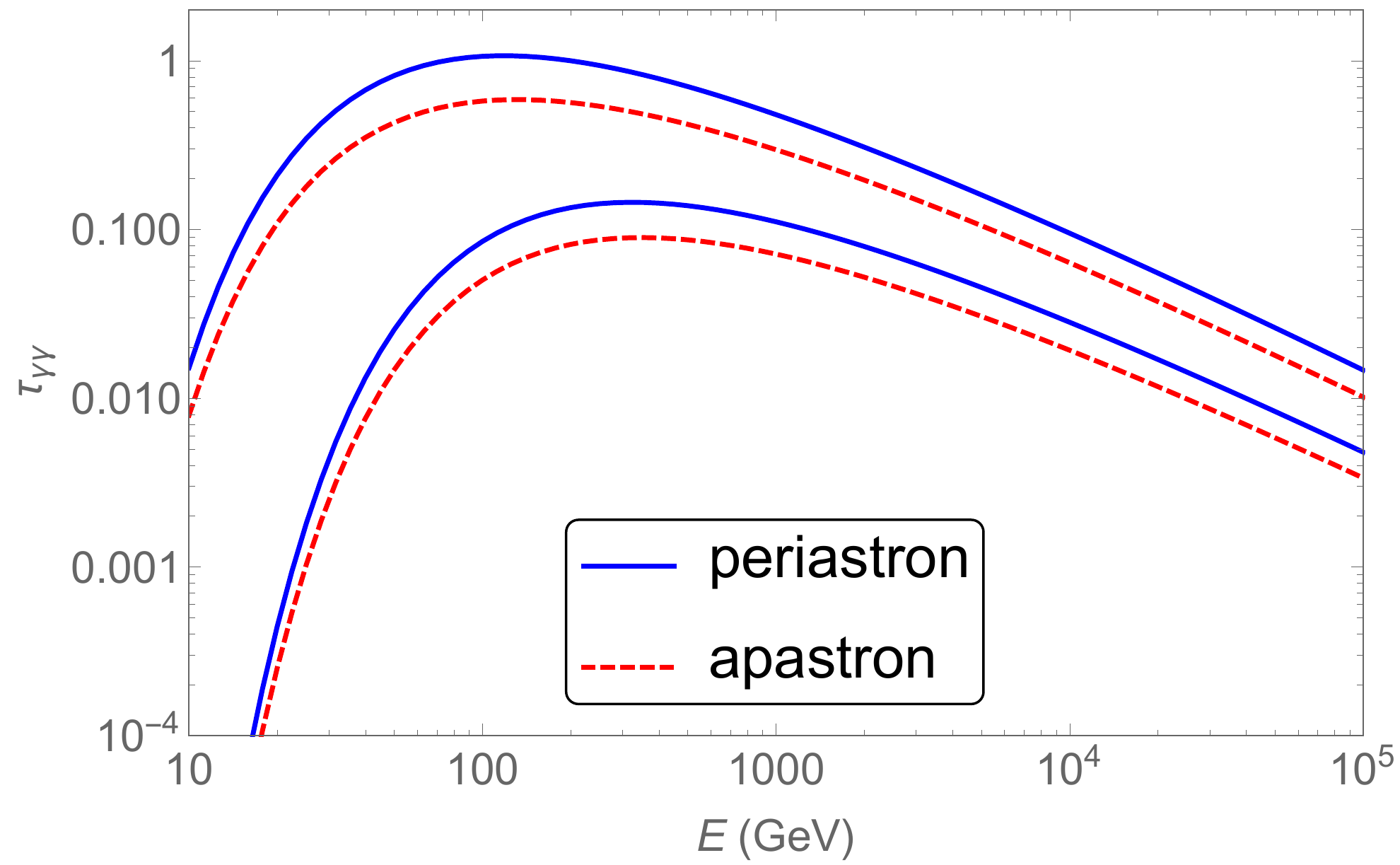}
\caption{$\gamma\gamma\to e^+e^-$ pair production opacity for gamma rays produced in the shocks of $\eta$ Carinae during the periastron and apastron passages.  The upper (lower) set of curves are for $\psi_0 = 0.25 \pi$ ($0.6\pi$).}
\label{Figure--opt}
\end{figure}

Gamma rays propagating at different angles $\psi_0$ will encounter different opacities.  Thus contributions to the observed flux from different parts of the shocked surface to an observer will vary.  The opacity is particularly high for emission from behind the stars~\cite{Ohm:2015uha}.  Here we have plotted the $\gamma\gamma$ opacity in Fig.~\ref{Figure--opt} for the periastron and apastron phases by integrating over $l$ from zero to the distance of $\eta$ Carinae.  We have set $\psi_0 = 0.6\pi$ (lower set of curves) as the direction from $\eta$ Car-A to the Earth~\cite{Pittard:1998} and $\psi_0 = \pi/4$ (upper set of curves) for illustrations. In general the opacity increases for decreasing $\psi_0$.  Based on these estimates, we infer that gamma-ray flux attenuation due to $\gamma\gamma$ pair production is not severe.  This is also consistent with the results found in~\cite{Ohm:2015uha} for 250~GeV gamma rays propagating along the direction of the Earth.  Future H.E.S.S. and CTA observations will be crucial to explore this further.  

\section{Neutrino flux and detection prospects}
We calculate the expected neutrino flux, generated by $pp\to \pi^\pm$ interactions and subsequent decays (see, e.g.,~\cite{Lunardini:2011br}), that should accompany $\pi^0$ decay gamma rays.  The proton spectrum given in Eq.~(\ref{pspecper}) with exponential cut-off at 1 PeV has been used to calculate the very high energy neutrino flux from $\eta$ Carinae. The neutrino flux for each flavor is assumed to be 2/3 of the gamma-ray flux~\cite{Gaisser:1990vg}.

\begin{figure}
\includegraphics[scale=0.35]{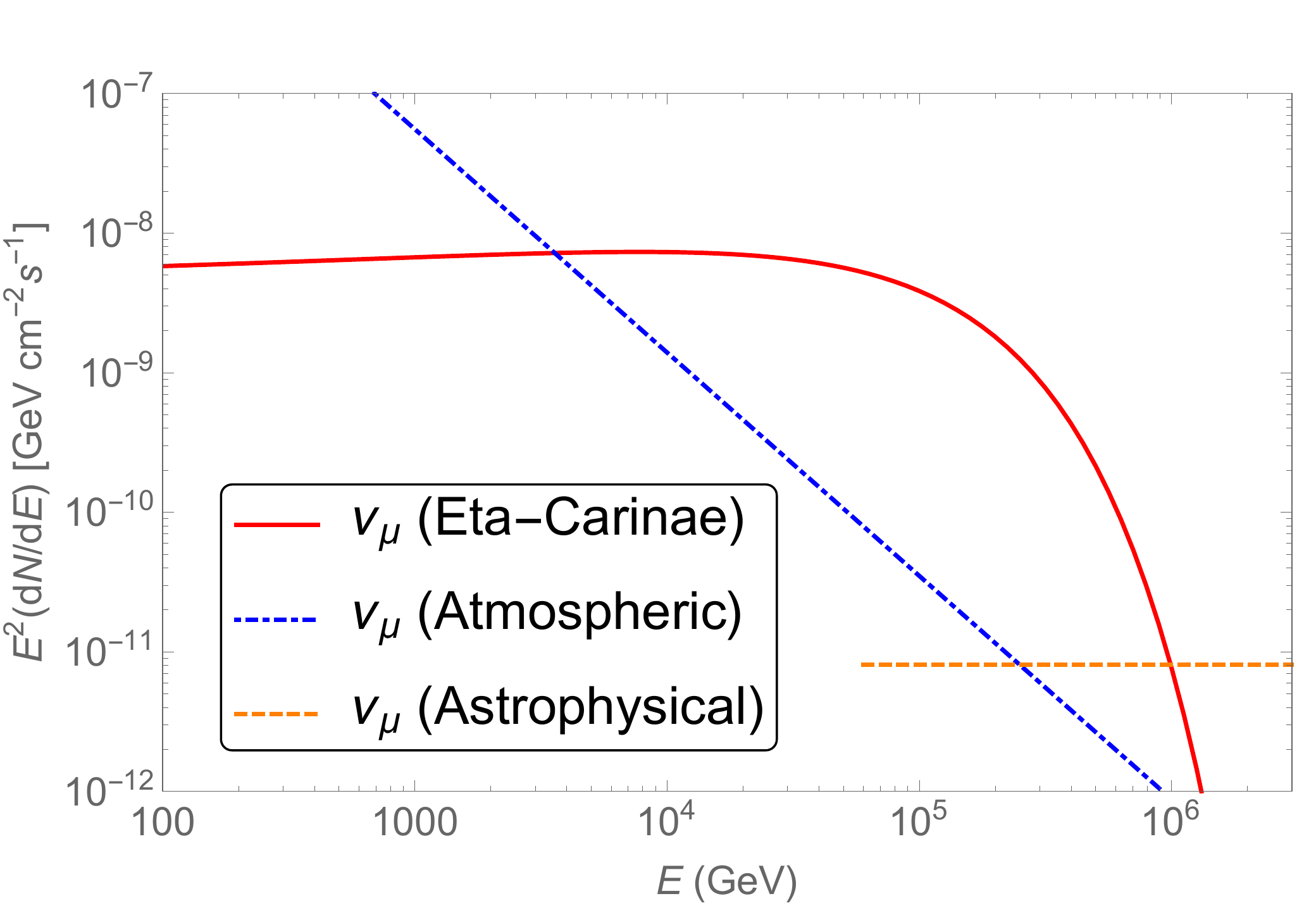}
\caption{Expected muon neutrino flux from $\eta$ Carinae during the 500-day periastron passage, corresponding to the gamma-ray flux shown in Fig.~\ref{Figure--1}.  Also shown are atmospheric neutrino flux and a diffuse astrophysical neutrino flux within $1^\circ$ of the direction of $\eta$ Carinae.}
\label{Figure--3}
\end{figure}

The muon-neutrino flux for the 500-day periastron passage is shown in Fig.~\ref{Figure--3} with the red solid line.  For neutrino telescopes, muon neutrinos provide the best angular resolution ($\sim 1^\circ$) for detection of any astrophysical object.  This corresponds to a solid angle $\Omega_\mu \approx 10^{-3}$~sr.  In Fig.~\ref{Figure--3} we have also plotted atmospheric muon-neutrino flux~\cite{Honda:2006qj} from within $1^\circ$ direction of $\eta$ Carinae (Right ascension: $10^{\rm h} 45^{\rm m} 03.591^{\rm s}$, Declination: $-59^\circ 41' 04.26''$) at the location of KM3NeT with blue dot-dashed line.  The astrophysical diffuse neutrino background detected by IceCube is given by $E^2\Phi(E) = (0.84\pm 0.3)\times 10^{-8}$~GeV~cm$^{-2}$~s$^{-1}$~sr$^{-1}$, with fixed spectral index~\cite{Aartsen:2015zva}.  This flux is also shown in Fig.~\ref{Figure--3}, after multiplying with $\Omega_\mu$, with orange dashed line.

It is clear from Fig.~\ref{Figure--3} that muon neutrino flux from $\eta$ Carinae dominates background at $\gtrsim 3$~TeV and can be a good target for neutrino telescopes.  For cascade events, corresponding mostly to electron and tau neutrinos, the angular resolution is much worse, $\sim 15^\circ$.  For these types of neutrino events, the atmospheric and astrophysical backgrounds would be a factor $\sim 220$ times higher, corresponding to the ratio of the solid angles.

\begin{figure}
\includegraphics[scale=0.37]{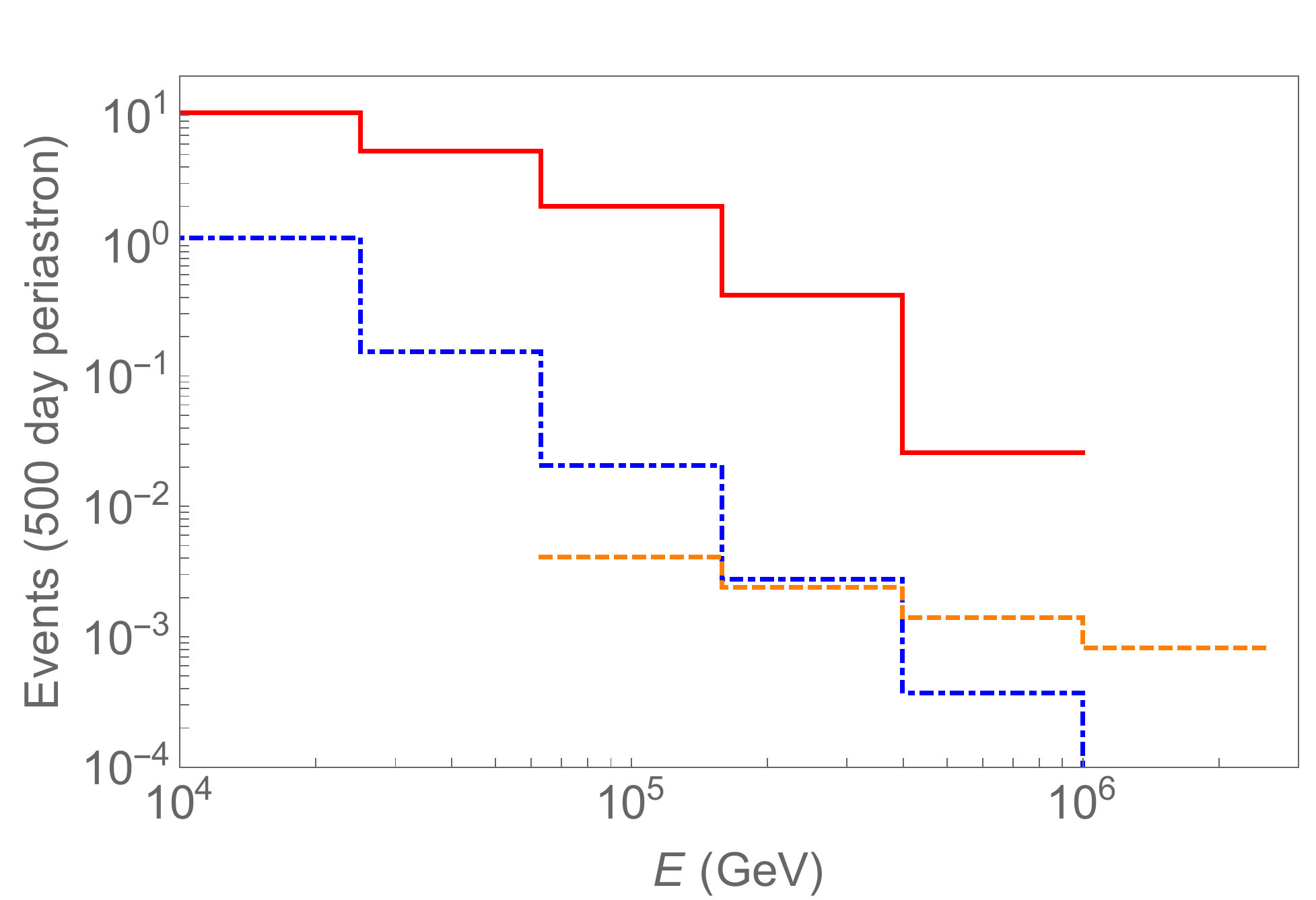}
\caption{Expected muon neutrino events from $\eta$ Carinae (red solid) in KM3NeT for an exposure equivalent to the 500-day periastron passage. Also shown are expected atmospheric (blue dot-dashed) and diffuse astrophysical (orange dashed) neutrino events from within $1^\circ$ of $\eta$ Carinae during the same period.}
\label{Figure--4}
\end{figure}

We calculate the number of muon-neutrino events in KM3NeT, with an exposure time of $T=500$ days, as (see, e.g., \cite{Lunardini:2011br, Lunardini:2013gva, Lunardini:2015laa})
\begin{equation}
N_{\nu_\mu} = T \int_{E_1}^{E_2} dE~\Phi(E) \langle A_{\nu_\mu} (E) \rangle,
\label{events}
\end{equation}
where $\Phi(E)$ corresponds to the fluxes plotted in Fig.~\ref{Figure--3} and $\langle A_{\nu_\mu} (E) \rangle$ is the average effective area for muon neutrinos in KM3NeT~\cite{Adrian-Martinez:2016fdl}.  Above 10~TeV the effective area can be approximated as $\langle A_{\nu_\mu} (E) \rangle \approx 10^4\,(E/{\rm GeV})^{0.42}\epsilon_d$~cm$^2$.  Here the pre-factor corresponds to the effective areas for all sky at trigger level (see figure 19 in~\cite{Adrian-Martinez:2016fdl}) and $\epsilon_d$ is an efficiency factor.  We note that $\eta$ Carinae has a similar Declination as the supernova remnant RXJ1713.7-3946, for which $\epsilon_d\approx 1$ at $E\gtrsim 10$~TeV assuming that all the muon track events are well reconstructed within $1^\circ$ from the source position (KM3NeT angular resolution is $\lesssim 0.2^\circ$-$0.3^\circ$ at $E\gtrsim 10$~TeV, see figures 22 and 25 in~\cite{Adrian-Martinez:2016fdl}) while their arrival directions are below the horizon (see figure 40 right panel in~\cite{Adrian-Martinez:2016fdl}). 

We show the distribution of muon-neutrino events in Fig.~\ref{Figure--4} above 10~TeV, where the energy range for each frequency bin for histograms correspond to the integration limits $E_1$ and $E_2$ in Eq.~(\ref{events}).  The total number of expected events between 10~TeV and 1~PeV from $\eta$ Carinae is 18.2 whereas the atmospheric background event is 1.3 and diffuse astrophysical event is 0.03 in the same energy range.  Assuming Gaussian statistics, we also calculate significance for detection of $\eta$ Carinae in neutrinos from event distributions in Fig.~\ref{Figure--3} as
\begin{equation}
\sigma =\sqrt{T\sum_i S_i^2/(S_i + B_i)} = 4.1\,,
\end{equation}
where $S_i$ and $B_i$ are events from $\eta$ Carinae and background (both atmospheric and astrophysical) in $i=1, ... 5$ energy bins.  Therefore, a $4\sigma$ detection would be possible during a 500 day periastron passage. KM3NeT is expected to have full visibility throughout the year to $\eta$ Carinae, located at a Declination $-59^\circ 41' 04.26''$ (see figure 37 in~\cite{Adrian-Martinez:2016fdl}). Therefore a lower significance detection can also be possible within a shorter time scale during the periastron passage.

Currently operating largest neutrino telescope IceCube is located at the South pole.  As a result, $\eta$ Carinae is above the horizon and atmospheric backgrounds are very high.  The High Energy Starting Event (HESE) selection reduces these backgrounds significantly but at the same time also reduces the effective area significantly.  The number of HESE neutrinos we calculate is 0.02 between 10~TeV and 1~PeV and for a 500-day periastron passage. 

In case the maximum proton energy is lower than 1~PeV that we have assumed, e.g., because of a smaller value of $\zeta_p$ in Eq.~(\ref{emax}), the maximum neutrino energy will be correspondingly lower.  In such a case neutrino detection will be less significant.

\section{Discussions and Conclusions}

We have modeled gamma-ray emission from the binary stellar system $\eta$ Carinae, particularly focusing on spectral energy distributions during the persiastron and apastron passages during its full orbit, from August 4, 2008 to February 18, 2014.  We found that a hadronic $\pi^0$ decay gamma-ray model from $pp$ interactions, which can describe $\gtrsim 10$~GeV data during the periastron passage also produces high-energy neutrinos.  We have modeled this neutrino flux and explored detectability of $\eta$ Carinae with neutrino telescopes.  We found that observations with KM3NeT during a 500-day periastron passage can lead to detection of $\gtrsim 10$~TeV neutrinos.  For IceCube, an improved effective area than HESE selection would be required for detection. {Observations of $\eta$ Carinae with neutrino telescopes is crucial for hadronic models of gamma-ray production and to constrain the maximum accelerated particle energy in stellar binaries.}

The SED of $\eta$ Carinae has been modeled by other authors earlier~\cite{Bednarek:2011yn, far, Ohm:2015uha}. The authors in~\cite{Bednarek:2011yn} have considered different acceleration and interaction scenarios, which would have a signature on the observed or lack of variability in the gamma-ray emission near 100 GeV. They have considered a pure leptonic model with two populations of electrons and also a lepto-hadronic model to explain the gamma-ray emission. It is hard to rule out any of these models with the GeV gamma-ray data alone. Their lepto-hadronic model could give gamma rays of energy 10 TeV-100 TeV. They have also predicted detectable neutrino event rate above the atmospheric background.

Our lepto-hadronic modeling is similar to~\cite{far}, however our work has been done with the full orbit data from August 4, 2008 to February 18, 2014.  They  have used 21 month Fermi LAT data in their analysis and predicted very low neutrino event rate for detection with KM3NeT in 5 years.

The authors of~\cite{Ohm:2015uha} found hadronic interactions more important than leptonic. They have used 5.5 years of full orbit data.  In their time-dependent model, gamma-ray absorption due to pair production is important above 100 GeV.  {We also found that $\gamma\gamma$ pair production with stellar photons from $\eta$ Car-A is important, although not severe.}  Future gamma-ray data at TeV energies would be useful for better understanding of the geometry of production region of gamma rays in this source.

Detection of high-energy neutrinos from $\eta$ Carinae would provide conclusive evidence of particle acceleration to PeV energies in colliding wind binaries as well as a hadronic contribution to the observed gamma-ray flux.  Our modeling can be extended to other stellar binary systems.  Predictable neutrino emission from these systems can be stacked for enhanced detection by the neutrino telescopes.

\section*{Acknowledgment}
We are thankful to R.~Coniglione for carefully reading the manuscript and providing detailed comments and to R.~Walter for helpful discussion.  We also thank K.~Reitberger for providing data and for helpful communications.  This work was supported by a grant from the University of Johannesburg (UJ) which allowed N.G.\ to travel to UJ where most of the work was carried out.

\end{document}